\documentclass[a4paper,american,reqno]{amsart}

\newif\ifPreprint \Preprinttrue
\newif\ifSubmission \Submissionfalse

\usepackage{nomencl}

\usepackage{babel}
\usepackage[utf8]{inputenc}
\usepackage[T1]{fontenc}
\usepackage{lmodern}
\usepackage[binary-units=true]{siunitx}
\usepackage{booktabs}
\usepackage{url}
\usepackage{xspace}
\usepackage{graphicx}
\usepackage{accents}
\usepackage{algorithm,algorithmic}
\usepackage{todonotes}
\usepackage{pgfplots}
\usepgfplotslibrary{groupplots}
\usepackage{subfigure}
\usepackage{tikz}
\usepackage{csquotes}
\usepackage{amssymb}
\usepackage{tabto}
\usepackage{siunitx}
\usepackage[official]{eurosym}
\usepackage[foot]{amsaddr}
\usepackage{microtype}
\usepackage{comment}
\usepackage[edges]{forest}
\usepackage[absolute,overlay]{textpos}
\usepackage{environ}
\makeatletter
\newsavebox{\measure@tikzpicture}
\NewEnviron{scaletikzpicturetowidth}[1]{%
  \def\tikz@width{#1}%
  \begin{lrbox}{\measure@tikzpicture}%
  \BODY
  \end{lrbox}%
  \pgfmathparse{#1/\wd\measure@tikzpicture}%
  \BODY
}
\makeatother
\usetikzlibrary{patterns}
\usepackage{pgfplots,pgfplotstable}
\usepackage[style = authoryear,
            maxbibnames = 100,
            maxcitenames = 2,
            isbn = false,
            backend = biber
            ]{biblatex}
\usepackage[colorlinks,
            citecolor=blue,
            urlcolor=blue,
            linkcolor=blue]{hyperref} % should always be the last package

\usepackage{amsmath}
\newtheorem{result}{Result}

\usepackage{etoolbox}
\renewcommand\nomgroup[1]{%
  \item[\bfseries
  \ifstrequal{#1}{A}{Sets and indices}{%
  \ifstrequal{#1}{B}{Parameters}{%
  \ifstrequal{#1}{C}{Variables}{}}}%
]}
\makeatletter
\patchcmd{\@settitle}{\uppercasenonmath\@title}{\scshape\large}{}{}
\patchcmd{\@setauthors}{\MakeUppercase}{\scshape\normalsize}{}{}
\makeatother

\vfuzz \hfuzz
\DeclareFieldFormat{eprint:urn}{%
  URN\addcolon\space
  \ifhyperref
    {\href{http://www.nbn-resolving.org/#1}{\nolinkurl{#1}}}
    {\nolinkurl{#1}}}

\DeclareFieldFormat{eprint:optonline}{%
  Optimization Online\addcolon\space
  \ifhyperref
    {\href{http://www.optimization-online.org/DB_HTML/#1.html}{\nolinkurl{#1}}}
    {\nolinkurl{#1}}}

\definecolor{light-gray}{gray}{0.85}
\definecolor{gray2}{gray}{0.68}
\definecolor{ultramarine}{RGB}{0,32,96}
\definecolor{darkblue}{RGB}{0, 183, 234}
\definecolor{verydarkblue}{RGB}{44, 127, 184}
\definecolor{lightblue}{RGB}{158, 202, 225}
\definecolor{green1}{RGB}{33,97,61}
\definecolor{green2}{RGB}{161,217,155}
\definecolor{blue1}{RGB}{141,160,203}
\definecolor{blue2}{RGB}{166,206,227}
\definecolor{blue3}{RGB}{31,120,180}
\definecolor{green3}{RGB}{178,223,138}
\definecolor{green4}{RGB}{141,211,199}

\definecolor{red}{RGB}{255,0,0}
\definecolor{blue}{RGB}{0,0,255}
\definecolor{green}{RGB}{0,255,0}
\definecolor{yellow}{RGB}{204,204,0}
\definecolor{violet}{RGB}{153,0,255}
\definecolor{orange}{RGB}{251, 208, 107}

\definecolor{green}{RGB}{0,160,0}

\bibliography{risk-aversion}

\begin{document}

\title[Risk aversion in flexible electricity markets]{Risk aversion in flexible electricity markets}
\author[Möbius, Riepin, Müsgens, van der Weijde]%
{Thomas Möbius$^{1}$, Iegor Riepin$^{1}$, Felix Müsgens$^{1}$, Adriaan H. van der Weijde$^{2,3}$}

\address{%
  $^1$ Brandenburg University of Technology Cottbus-Senftenberg, Siemens-Halkse-Ring 13, 03046 Cottbus, Germany;
  $^2$ The University of Edinburgh, School of Engineering, Robert Stevenson Road, The King's Buildings, Edinburgh EH9 3FB, UK;
  $^3$ The Alan Turing Institute, 96 Euston Rd, London NW1 2DB, UK.
}

\email{$^1$thomas.moebius@b-tu.de}
\email{$^2$iegor.riepin@b-tu.de}
\email{$^3$felix.muesgens@b-tu.de}
\email{$^4$H.Vanderweijde@ed.ac.uk}

\date{\today}

\begin{abstract}
  Flexibility options, such as demand response, energy storage and interconnection, have the potential to reduce variation in electricity prices between different future scenarios, therefore reducing investment risk. Moreover, investment in flexibility options can lower the need for generation capacity. However, there are complex interactions between different flexibility options. In this paper, we investigate the interactions between flexibility and investment risk in electricity markets. We employ a large-scale stochastic transmission and generation expansion model of the European electricity system. Using this model, we first investigate the effect of risk aversion on the investment decisions. We find that the interplay of parameters leads to (i) more investment in a less emission-intensive energy system if planners are risk averse (hedging against CO\textsubscript{2} price uncertainty) and (ii) constant total installed capacity, regardless of the level of risk aversion (planners do not hedge against demand and RES deployment uncertainties). Second, we investigate the individual effects of three flexibility elements on optimal investment levels under different levels of risk aversion: demand response, investment in additional interconnection capacity and investment in additional energy storage. We find that that flexible technologies have a higher value for risk-averse decision-makers, although the effects are nonlinear. Finally, we investigate the interactions between the flexibility elements. We find that risk-averse decision-makers show a strong preference for transmission grid expansion once flexibility is available at low cost levels.

  \smallskip
  

  
\end{abstract}

\maketitle
\section{Introduction}
\label{sec:introduction}
 Investments in electricity transmission and generation capacity are made under a significant amount of uncertainty. The sources of this uncertainty include, among others, the future levels and spatio-temporal distributions of electricity demand, fuel costs and future energy policy. In recent years, these uncertainties have arguably increased due to the ongoing evolution of supply- and demand-side technologies and rapid policy changes designed to encourage a transition to low-carbon energy systems. Because transmission and generation investments have long lead times and are difficult to reverse, they are subject to a large -- and arguably increasing -- amount of risk. 
 
As in most markets, investors in electricity markets are usually risk averse (\cite{wustenhagen2006, salm2018,ostrovnaya2020}. This risk aversion arises from the risk aversion of individual decision-making agents within firms, the use of risk mark-ups on the expected return on investment, %hurdle rates
the use of risk-averse probabilistic metrics by the institutions that finance new projects, and other factors. There is evidence that, separately from the increase in energy investment risk, investor risk aversion itself is increasing as a result of the participation of less traditional actors, such as institutional investors, in energy markets (\cite{salm2018}). Therefore, the literature on transmission and generation planning has begun to include risk-averse objectives in planning models (e.g. \cite{fan_et_al:2010, ehrenmann_smeers:2011, delgado_claro:2013, Munoz2017, ambrosius_et_al:2019}. Although the exact impacts of risk aversion differ according to the specifics of the settings analysed and the types of uncertainty considered, most of these studies show that risk aversion significantly changes optimal transmission and generation expansion plans. 

These insights are useful. However, the existing literature has several gaps that we aim to address in this paper. Most importantly, previous studies have tended to analyse current electricity markets, which feature only limited deployment of flexibility elements, such as energy storage, demand response and international integration. In all likelihood, future electricity markets will significantly increase the deployment of these flexibility elements. This is especially important in the context of risk aversion. These sources of flexibility will enable the more efficient integration of low-carbon generation technologies, allowing fuel price and carbon price risk to be addressed at lower cost. They can also reduce demand-side risk by decreasing demand peaks. Finally, flexible technologies can reduce spatial risks by allowing electricity to be transported more efficiently. Therefore, the risk profiles of future electricity markets are likely to differ substantially from those of current markets; by extension, the effects of risk aversion will change as well.

A second, closely related gap is that the existing literature not only tends to analyse relatively inflexible electricity markets, it also generally does not explicitly include investment in flexible technologies such as storage or interconnection. In addition to their different risk profiles, flexible systems offer a wider set of investments that can be undertaken to mitigate risk exposure. These may include investment decisions in flexible technologies in addition to generation capacity. 
 
Third, the existing literature on risk aversion includes renewable investment as an endogenous variable. Since renewable investment is an effective hedge against fuel and carbon price risk, as has been shown in the literature (e.g. \cite{Munoz2017}), it often drives model results; suggesting a higher amount of investment in renewables when decision-makers are risk averse. In many jurisdictions, however, renewable investment is driven more by policy and central planning than by market forces. Therefore, such investment might not serve as a relatively inexpensive hedge against high fuel or carbon prices. 
 
In this paper, we explicitly analyse the interactions between risk aversion and flexibility. Since international interconnection is one of the most widely recognised elements of flexibility, we cannot simply focus on a single electricity market but instead consider a series of linked markets. This simultaneously addresses a fourth gap in the existing literature on risk aversion, which typically focuses on individual markets even though investors often operate across markets. Our case study focuses on the European electricity market, which consists of several linked but independent national markets. Using an integrated planning model applied to a simplified representation of the European electricity grid, we consider three types of flexible technologies: increasing interconnection between national markets, energy storage investment and demand flexibility. As mentioned above, increasing interconnection provides flexibility because it allows the movement of electricity -- especially low-cost renewable electricity -- in space. Energy storage can move electricity in time. Finally, demand flexibility allows consumers to more efficiently signal their valuation of electricity, allowing scarce resources to be allocated more efficiently and reducing the need for investment in peak-load capacity with low utilisation rates.
 
We analyse the impact of each of these sources of flexibility on the system in turn and then also consider the interactions between them. Of particular importance are the interactions between increasing interconnection and demand flexibility. Since each country has its own distinct industry structure, with different levels of willingness to pay between industries and between countries, increasing flexibility can increase overall efficiency but also significantly affect the spatial distribution of investment and net demand. Unlike earlier studies, we consider renewable investment to be exogenous (i.e. determined by existing long-term national renewable expansion plans rather than by electricity market prices) but stochastic (i.e. there is uncertainty about how successful current plans will be). We also consider demand, fuel price and carbon price uncertainty using an inter- and extrapolation of the widely used ENTSOs Ten-Year Network Development Plan (TYNDP) scenarios (\cite{tyndp2018}).
 
This approach allows us to address the four gaps identified above. In addition to contributing to the literature on risk-averse planning by generalising it to flexible energy systems, we also advance the literature on the value of flexibility elements by including the value of energy storage, interconnection and demand response. We generalise this literature by including the value of flexibility in terms of risk reduction, which, as we will see, is important if decision-makers are risk averse. 
 
In line with the previous literature, we find that risk aversion significantly affects investments in transmission and generation capacity. Although we include uncertainty in demand, renewable capacity and fuel prices, we find that carbon price uncertainty generates the highest level of investment risk in the European context. In response, risk aversion decreases investment incentives especially for high-emission technologies despite the risk of failing to meet electricity demand. Conversely, risk aversion increases investment in international interconnection capacity when this option is available. Risk aversion also increases incentives to invest in energy storage and to utilise demand flexibility. Finally, there are important nonlinear interactions between these flexibility elements. In particular, when all flexibility options are available at low cost, risk-averse system planners increasingly use the interconnection expansion option, reducing the reliance on demand flexibility and storage expansion. When flexibility options are available but more expensive, the opposite happens.

The next section briefly reviews the existing literature on risk aversion in electricity markets. Section \ref{sec:methods} sets out the details of our modelling methodology. Section \ref{sec:scenarios and data} provides all our assumptions regarding the scenarios and data. Our results are presented in Section \ref{sec:results} and discussed in Section \ref{sec:discussion}. Section \ref{sec:conclusion} concludes.

\section{Literature review}
\label{sec:literature}

Questions concerning optimal or expected generation and transmission capacity expansions have been a subject of research for decades. A wide range of models have been developed to help determine optimal transmission expansion plans, simulate prices and investment incentives under different future scenarios, determine the impact of market design changes, among others. Early models were deterministic and often tailored towards either generation or transmission expansion in isolation. Currently, the most common approach integrates transmission and generation expansion into a multi-level partial equilibrium model of the electricity sector. If markets are assumed to be perfectly competitive, which is the usual approach, and transmission planners are assumed to maximise social surplus, the problem can be condensed into a single-level optimisation model (\cite{garces2009}). 

One key finding that has been shown repeatedly is that explicitly considering uncertainty is important when modelling generation and transmission expansion. Models that rely on the heuristic application of deterministic models lead to a significant cost of ignoring uncertainty, as shown by \textcite{RIEPIN2021}, who investigate the expected costs of ignoring uncertainty for investments in an integrated electricity and gas market model. In response to this, and to rapid decreases in the computational cost of large-scale models, stochastic capacity expansion models have become more common (e.g. \cite{van_der_weijde_economics_2012,Pozo2013,ambrosius_et_al:2019}).

Most stochastic studies use simple expected-value optimisation models. While such models are useful, they do not account for the fact that real-world investors are risk averse and therefore under- or overestimate investment incentives for particular technologies or locations. In response, risk-averse objectives have been included using either nonlinear utility functions (e.g. \cite{fan_et_al:2010}) or, more recently, coherent risk metrics (e.g. \cite{Munoz2017}). These studies show that, like uncertainty, risk aversion changes investment strategies. Generally, higher levels of risk aversion lead to more investment in renewables (\cite{TIETJEN2016174,INZUNZA2016104,Diaz2019}) complemented by lower-cost thermal capacity (\cite{ehrenmann_smeers:2011}), although the precise implications depend heavily on the types and relative magnitudes of the uncertainties considered.

At the same time, scholars have shown increasing interest in considering investment options other than straightforward transmission and generation capacity. Studies have examined storage investment (e.g. \cite{fernandez2017, BOFFINO2019}), investment in the presence of demand response (\cite{DEVINE2019}) investment in power plant flexibility (\cite{GARDARSDOTTIR2018}), market-participating residential microgeneration (\cite{CALVILLO2016}), as well as other demand-side flexibility options (\cite{moreno2017}).

So far, however, little attention has been paid to the interactions between risk aversion and flexibility in electricity markets. These interactions are clearly important. \textcite{Diaz2019} use a risk-averse stochastic planning model to show that electricity storage reduces price risk and is therefore more attractive as part of a risk-averse portfolio. The same is likely to be true for other flexibility options. There are also obvious interactions between different types of flexibility. For instance, \textcite{Cepada2009} use a simple statistical model to demonstrate that interconnection and load shedding, which can be thought of as a last-resort type of demand flexibility, are substitutes. \textcite{Cepada2009} also show that in connected asymmetric markets, the effects of adding flexibility options will also be asymmetric, with some markets benefiting more than others. 

Therefore, there is a gap in the existing literature, which we aim to address in this paper. We consider the interactions between risk aversion and three types of flexibility: demand response, international interconnection and energy storage. As we are interested in empirical results in addition to theoretical effects, we employ a large stochastic model of the European electricity system combined with a widely used set of scenarios and a detailed representation of demand flexibility. Instead of the common approach, which uses continuous direct demand functions which are often difficult to parameterise, we approximate the real-world flexibility of demand by constructing piecewise linear demand functions based on the different components of electricity demand in each market and their estimated value of lack of adequacy (VoLA\footnote{\textcite{CEPA:2018} defines this as the equivalent of the value of lost load (VoLL) when notice is provided one day ahead of the supply interruption}). We exogenously vary the amount of these flexible options to determine their individual and joint effects, but also consider cases where investment in storage and interconnection is endogenous.

\section{Methods}
\label{sec:methods}

In this section, we first present a general form of the stochastic optimisation problem that we apply. Second, we introduce our approach to modelling the risk aversion of market participants. Third, we explain how we translate this approach to a large-scale optimisation model of the European electricity market. Fourth, we discuss the input data for the model.

\subsection{Two-stage stochastic optimisation}
\label{subsec:methods_section_1}

Consider a linear optimisation problem (LP) where $x$ represents a vector of variables, $c$ and $b$ are vectors of parameters, $A$ is a matrix of parameters and $T$ is a matrix transpose. The inequalities specify a convex polytope that constrains the objective function. The optimal deterministic solution (DS) is to find a vector $x$ that minimises the objective function given a set of constraints:

\begin{equation}
\label{eqn:eq1}
\begin{array}{l}
DS:= \displaystyle\min_{x} c^Tx \\
\textrm{s.t. } Ax \geq b, x \geq 0
\end{array}
\end{equation}
\vspace*{5px}

Deterministic linear problems of this type can be scaled to large optimisation problems with high levels of empirically relevant detail for the energy sector. They can then be solved with efficient linear solvers such as CPLEX and Gurobi. 

However, whenever energy planners attempt to construct a model of a real-world electricity market, they are faced with multiple parametric uncertainties. This is true irrespective of the application domain. These uncertainties include (market-driven) prices for energy carriers, the future development of demand levels and CO\textsubscript{2} prices, structural changes in the energy sector (e.g. nuclear phase-outs), regulation (e.g. decolonisation policies and the subsidisation of renewable generation) and many more factors.

In the literature, this parametric uncertainty is typically addressed \emph{reactively} -- through sensitivity analyses -- or \emph{proactively} -- through stochastic programming formulations (\cite{Mulvey1995}). The aim of the first approach is to reveal the impact of data perturbations on a model’s recommendations. Such an exercise is \emph{reactive} in that it only discovers the impact of data uncertainty on model outputs and requires a predefined heuristic rule to make a decision that is \emph{good enough} given all the scenarios under consideration. 

The aim of the (\emph{proactive}) approach is to explicitly incorporate uncertainty as part of the optimisation problem. This can be done by formulating a two-stage stochastic linear optimization problem that uses a probability distribution or scenario set to describe the uncertain parameters. In such a problem, the optimal \emph{first-stage} decisions must be made before the information about uncertain parameters is revealed, while the \emph{second-stage} decisions are made after it is revealed. The general form of a two-stage stochastic LP as described in \textcite{birge_louveaux:2011} is as follows:

\begin{equation}
\label{eqn:eq2}
\begin{array}{l}
SS:= \displaystyle\min_{x} c^Tx + \mathbb{E}_{s}[Q(x,s)] \\
\textrm{s.t. } Ax \geq b, x \geq 0 \\
\end{array}
\end{equation}

\begin{equation}
\label{eqn:eq3}
\begin{array}{l}
\textrm{where} \\
Q(x,s) := \displaystyle\min_{y} q^T_{s}y \\
\textrm{s.t. } T_{s}x + W_{s}y \geq d_{s}, y \geq 0 \\
\end{array}
\end{equation}
\vspace*{5px}

where $x$ denotes the vector of first-stage variables, $y$ is the vector of second-stage variables and $s$ is the vector of uncertain data (i.e. the discrete set of scenarios that include a finite number of scenarios with differences regarding the uncertain parameters). Each scenario $s \in S$ occurs with a probability $\sum_s \pi_{s} = 1$. The first-stage parameters ($c^T, A, b$) in Eq. (\ref{eqn:eq2}) are assumed to be known with certainty. The second-stage decisions in Eq. (\ref{eqn:eq3}) are restricted by the first-stage decisions $x$, and the parameters ($q,T,W,d$) are actual realisations of uncertain data (i.e. the solution must be feasible for every scenario $s \in S$). Overall, the first-stage problem seeks to minimise the sum of the costs of the first-stage decisions and the \emph{expected} costs of the second-stage decisions. The second-stage problem accordingly seeks to minimise the second-stage costs.

In the context of our electricity market application, solving problem (\ref{eqn:eq2}--\ref{eqn:eq3}) implies finding both (i) the optimal \emph{first-stage} investment decisions for power generation and transmission capacities $\hat{x}$  under uncertainty by considering the probability of the uncertain events\footnote{Note that investment decisions must hold for \emph{all realisations} of $s \in S$} and (ii) the optimal \emph{second-stage} dispatch decisions (generation, storage and demand response) for \emph{each realisation} of $s \in S$ (i.e. after market agents have full knowledge of the uncertain parameters).

\subsection{Modelling risk aversion}
\label{subsec:methods_section_2}

As discussed above, individual decision-making agents in the electricity markets are usually risk averse. Hence, we include risk aversion in the objectives of the mathematical models when modelling investment decisions and system development. In general, this involves replacing the expectation operator in \ref{eqn:eq2} with a risk measure. In line with the previous literature (e.g. \cite{Munoz2017}), we use a convex combination of the second-stage expected value and the conditional value at risk (CVaR) (\cite{rockafellar_et_al:2000}). The CVaR is a coherent risk measure that effectively increases the weight on the $1-\alpha\%$ worst outcomes, where `worst' is defined endogenously. Since it is the solution to a linear optimisation problem, it can easily be integrated into the stochastic framework described above. Increasing the weight of a discrete set of worst outcomes is also consistent with the use of P90 percentiles and threshold values in energy finance (\cite{MORA2019}), which is one reason why investors are often risk averse.

\subsection{The stochastic optimization problem for European electricity markets with risk aversion}
\label{subsec:methods_section_3}

Following the model structure outlined above, we develop and apply a stochastic large-scale optimisation model of the European electricity sector considering risk-averse decision-makers. The model results include spatial and temporal decisions about investments in electricity generation, transmission and storage capacities, as well as decisions about the generation, trade and storage of electricity. 

The geographic scope of the model includes most EU27 member states\footnote{Cyprus, Ireland, Iceland and Malta are not included}, Norway, Switzerland and the United Kingdom. The single market zones are connected through net transfer capacities (NTCs). A list of the regions considered is presented in Appendix B. The temporal scope of the model includes three representative years: 2020, 2025 and 2030; each year is modelled by 350 representative operational hours. This approach ensures the inclusion of both long-term investment dynamics and the short-term dispatch of infrastructure elements. Investment decisions are made once, at the beginning of the modelled period.

In the following, we present our optimisation model. We first establish a nomenclature and then provide all relevant model formulations.

\subsubsection{Nomenclature}
% Nomenclature formatting in backmatter.tex
% Documentation: https://www.overleaf.com/learn/latex/Nomenclatures

Note that we use capital letters for all sets and parameters except for the parameters $\alpha$, $\eta_{i}$, $\rho_{s}$ and $\omega$.

\begin{figure}[H]
\includegraphics[width=\textwidth]{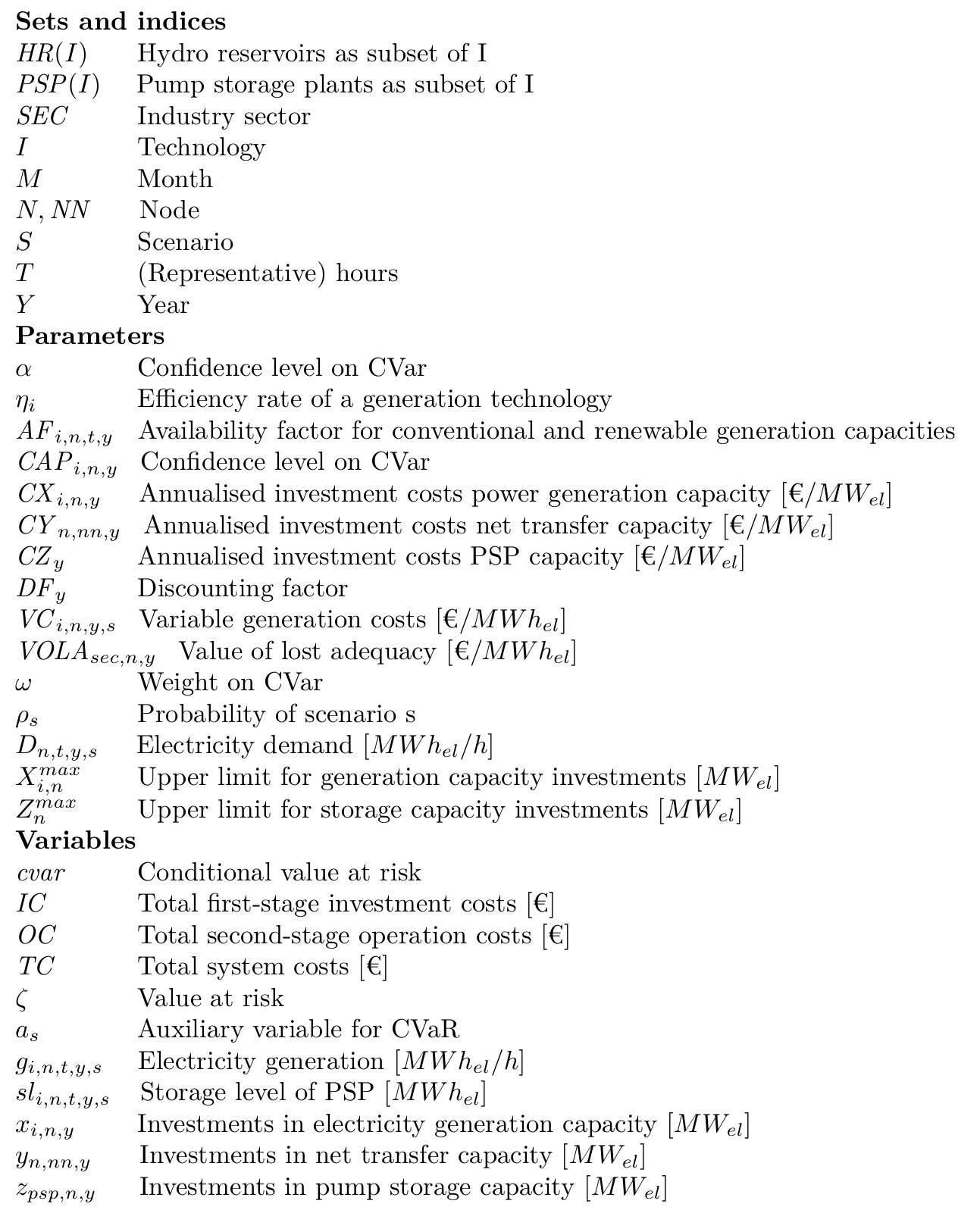}
\end{figure}

\subsubsection{Model formulation}

The objective function in Eq. \ref{eqn:obj} minimises total system costs considering the costs of all first-stage investments $IC$ and second-stage operation costs $OC$. With increasing risk aversion, decision-makers give a higher weight $\omega$ to the riskiest scenarios considered in $cvar$. 

\begin{equation}
\label{eqn:obj}
\begin{aligned}
    \displaystyle\min TC= IC + (1-\omega) \sum_{s} \rho_s \mathit{OC}_s + \omega \mathit{cvar}\\
\end{aligned}
\end{equation}

The investment costs $IC$ include all discounted annualised investments in power generation capacities $x_{i,n,y}$, cross-border transmission capacities $y_{n,nn,y}$ and pumped-storage plant (PSP) capacities $z_{psp,n,y}$.

\begin{equation}
\label{eqn:IC}
\begin{aligned}
    \displaystyle IC = \sum_{y} DF_{y} \left( \sum_{i,n} CX_{i,y} x_{i,n,y} + \sum_{n} CY_{n,nn,y} y_{n,nn,y} + \sum_{n} CZ_{y} z_{psp,n,y} \right)\\
\end{aligned}
\end{equation}

Note that the investments made are irreversible. Therefore, the annuities for the investments made at the beginning of the model period must be taken into account in all subsequent years.

\begin{equation}
\label{eqn:IC_restriction_min}
\begin{aligned}
    \displaystyle   x_{i,n,y}-x_{i,n,y+1},  y_{n,nn,y}-y_{n,nn,y+1},  z_{psp,n,y}-z_{n,y+1} \leq 0  \\
    \forall i\in I, n, nn\in N, y\in Y 
\end{aligned}
\end{equation}

Eqs. \ref{eqn:IC_restriction_max_1} and \ref{eqn:IC_restriction_max_2} take into account the country-specific technical and political restrictions that limit investments in power generation and storage capacity. For example, countries with a nuclear phase-out strategy cannot invest in nuclear power plants.
\begin{equation}
\label{eqn:IC_restriction_max_1}
\begin{array}{l l}
      x_{i,n,y} \leq X^{max}_{i,n} & \forall i\in I,n\in N,y\in Y \\
\end{array}
\end{equation}

\begin{equation}
\label{eqn:IC_restriction_max_2}
\begin{array}{l l}
      z_{psp,n,y} \leq Z^{max}_{n} & \forall n\in N,y\in Y \\
\end{array}
\end{equation}

In the second stage, the operational decisions are made depending on the scenario realisation $s$. Therefore, the second-stage operational costs $OC$ include discounted costs of electricity production and demand response.

\begin{equation}
\label{eqn:OC}
\begin{aligned}
    \displaystyle OC_s = \sum_{y}DF_y \left(\sum_{i,n,t} VC_{i,n,y,s} g_{i,n,t,y,s} + \sum_{sec,n,t} VOLA_{sec,n,y} shed_{sec,n,t,y,s}\right) \\
    \forall s\in S  \\
\end{aligned}
\end{equation}

To define the CVaR, we apply the approach used in \cite{ Munoz2017}.
\begin{equation}
\label{eqn:Cvar_1}
\begin{aligned}
    \displaystyle \zeta + \frac{1}{1-\alpha} \sum_{s\epsilon S} \rho_s a_s \leq cvar\\
\end{aligned}
\end{equation}

\begin{equation}
\label{eqn:Cvar_2}
\begin{array}{l l}
    \displaystyle a_s \geq OC_s - \zeta & \forall s\in S\\
\end{array}
\end{equation}

Eq. \ref{eqn:energy_balance} ensures efficient market clearing, where demand $D_{n,t,y,s}$ always equals the sum of generation $g_{i,n,t,y,s}$ of all conventional, renewable and storage technologies, increased by imports $flow_{nn,n,t,y,s}$ and reduced by exports $flow_{n,nn,t,y,s}$ and the energy withdrawal of storage facilities $pump_{psp,n,t,y,s}$.

\begin{equation}
\label{eqn:energy_balance}
\begin{aligned}
    \displaystyle D_{n,t,y,s} = \sum_{i}g_{i,n,t,y,s} - \sum_{psp}pump_{psp,n,t,y,s}  + \sum_{nn} (flow_{nn,n,t,y,s} \\ 
    - flow_{n,nn,t,y,s} )  + \sum_{sec}shed_{sec,n,t,y,s} \hspace{1em}  \forall n\in N,t\in T,y\in Y,s\in S  \\
\end{aligned}
\end{equation}

Electricity generation is constrained by the existing capacities $\mathit{CAP}_{i,n,y}$, new investment $x_{i,n,y}$ and availability factors $AF_{i,n,t,y}$. Note that neither intermittent nor dispatchable renewables are subject to investment decisions since the expansion path for these technologies is implemented exogenously and is subject to uncertainty. Curtailment occurs in cases where renewable generation is lower than available capacity. Penalty payments for renewable curtailment are not considered.

\begin{equation}
\label{eqn:max_Generation}
\begin{array}{l l}
    \displaystyle g_{i,n,t,y,s} \leq (CAP_{i,n,y} + x_{i,n,y}) AF_{i,n,t,y} & \forall i\in I, n\in N ,t\in T, y\in Y, s\in S \\ 
\end{array}
\end{equation}

We model demand response through a supply function derived from sector-specific values of lost adequacy (VoLAs), as explained in more detail below. Therefore, hourly sector-specific demand reductions in each sector are limited to the sectoral electricity consumption in the respective hour. 
\begin{equation}
\label{eqn:shed_max}
\begin{aligned}
    \displaystyle shed_{sec,n,t,y,s} \leq D_{n,t,y,s} SHARE_{sec,n} \\
    \forall sec\in SEC, n\in N, t\in T, y\in Y, s\in S \\ 
\end{aligned}
\end{equation}

Our model also considers energy storage. For this purpose, we exclusively consider PSPs. The hourly pumping capacity of PSPs $pump_{psp,n,t,y,s}$ is restricted by their existing and newly built capacity. Note that turbine generation is already addressed in Eq. \ref{eqn:max_Generation}.
\begin{equation}
\label{eqn:Pump_max}
\begin{aligned}
    \displaystyle  pump_{psp,n,t,y,s} \leq (CAP_{psp,n,y} + z_{psp,n,y}) AF_{psp,n,t,y} \\
    \forall psp\in I, n\in N, t\in T, y\in Y, s\in S \\ 
\end{aligned}
\end{equation}

The storage level $sl_{psp,n,t,y,s}$ defines the water level in the upper basin of a PSP. It increases due to pumping activities and decreases due to the PSP's electricity generation. Efficiency losses $\eta_{psp}$ are also considered.
\begin{equation}
\label{eqn:storage_level}
\begin{aligned}
    \displaystyle  sl_{psp,n,t,y,s} = sl_{psp,n,t-1,y,s} + pump_{psp,n,t,y,s} \eta_{psp} - g_{psp,n,t,y,s} \\
    \forall psp\in I, n\in N, t\in T, y\in Y, s\in S \\ 
\end{aligned}
\end{equation}

The storage level is restricted to the maximum capacity of the upper basin of a PSP, which is linked to the installed generation capacity and a capacity--power factor. The capacity--power factor can be understood as the number of full load hours a PSP can run until it is fully discharged.  
\begin{equation}
\label{eqn:storage_level_max}
\begin{aligned}
    \displaystyle  sl_{psp,n,t,y,s} \leq (CAP_{psp,n,y} + z_{psp,n,y}) CPF \\
    \forall psp\in I, n\in N, t\in T, y\in Y, s\in S  \\ 
\end{aligned}
\end{equation}

Eq. \ref{eqn:hydro_reservoir} introduces a monthly water budget for hydro reservoirs that cannot be exceeded due to the electricity generation of these power stations.
\begin{equation}
\label{eqn:hydro_reservoir}
\begin{aligned}
    \displaystyle  \sum_{t\epsilon m} g_{hr,n,t,y,s} \leq CAP_{hr,n,y} budget_{hr,n,m} \\
    \forall hr\in I, n\in N, m\in M, y\in Y, s\in S  \\ 
\end{aligned}
\end{equation}

The cross-border flow of electricity is constrained by the existing NTCs $NTC_{n,nn,y}$ and newly built transfer capacities $y_{n,nn,y}$. 
\begin{equation}
\label{eqn:line_flow}
\begin{aligned}
    \displaystyle flow_{n,nn,t,y,s} \leq NTC_{n,nn,y} + y_{n,nn,y} \\
    \forall n\in N, nn\in N, t\in T, y\in Y, s\in S   \\ 
\end{aligned}
\end{equation}

Non-negativity:
\begin{equation}
\label{eqn:Nonnegativity}
\begin{aligned}
  \displaystyle    a_s,  g_{i,n,t,y,s}, flow_{n,nn,t,y,s}, pump_{psp,n,t,y,s}, shed_{sec,n,t,y,s}, sl_{i,n,t,y,s},  x_{i,n,y}, \\ 
  y_{n,nn,y}, z_{psp,n,y},  \geq 0 
      \hspace{1em}      \forall  i\in I, sec \in SEC, n,nn\in N, t\in T, y\in Y, s\in S \\
\end{aligned}
\end{equation}

\section{Scenarios and Data}
\label{sec:scenarios and data}

In this section, we explain the development of our scenarios and provide the assumptions for the data used in our model.

\subsection{Scenarios}
\label{subsec:Scenarios}

We parametrise our model primarily based on the widely used Ten-Year Network Development Plan (TYNDP) report published by ENTO-E (\cite{tyndp2018}). For the year 2030, the TYNDP presents three scenarios describing possible energy futures: distributed generation (DG), sustainable transition (ST) and the European Commission’s core policy scenario (EUCO). 

In the context of these scenarios, we identify four key parameters in the electricity sector that are subject to uncertainty: electricity demand, installed intermittent renewable energy source (RES) capacities (such as onshore wind, offshore wind and photovoltaics), fuel prices and CO\textsubscript{2} prices. In the following, we briefly describe the scenarios and highlight their assumptions.

The DG scenario assumes the development of a decentralised energy system with a strong focus on end-user technologies. As a result, electric vehicles and hybrid heat pumps see their highest penetration in this scenario, and photovoltaics and battery systems are widely used. Of the three scenarios, DG assumes the highest levels of electricity consumption (both peak and cumulative) and the highest deployment of solar power. 

The ST scenario assumes very high CO\textsubscript{2} prices intended to achieve a sustainable reduction of CO\textsubscript{2}, resulting in the replacement of hard coal and lignite by gas-fired power plants. The electrification of other sectors, such as heat and transportation, develops rather slowly. In comparison to the other scenarios, ST shows the highest price for CO\textsubscript{2} certificates (89.9 €2020/t in 2030) and the lowest electricity demand.

Finally, EUCO represents the 2016 EU Reference Scenario. The scenario represents a pathway to achieve the 2030 climate and energy targets agreed upon by the European Council in 2014. Compared to the other scenarios, EUCO shows the highest prices for lignite and hard coal and the lowest prices for natural gas and oil. Additionally, EUCO assumes the lowest price for CO\textsubscript{2} certificates (28.8 €2020/t in 2030).

\begin{figure}[H]
\includegraphics[width=\textwidth]{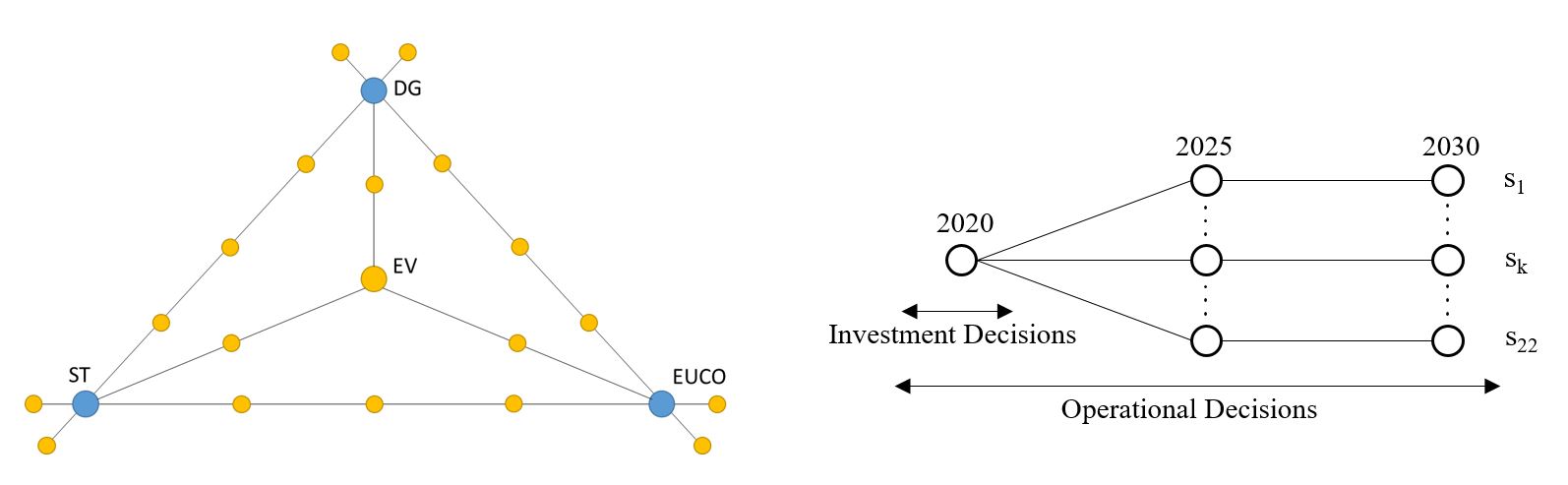}
	\caption{Scenario generation (left) and scenario paths (right) based on TYNDP scenarios.}
    	\label{figure_scenarios}
\end{figure}

Since we use the CVaR to analyse the effect of risk aversion, we must consider more than three scenarios to meaningfully differentiate a small percentage of tail scenarios from the others. We address this requirement by generating a range of scenarios that are linear inter- and extrapolations of the TYNDP scenarios. We also compute the expected value (EV), which we calculate as the average of the three TYNDP scenarios. This allows us to generate three additional scenarios by interpolating between the EV and each TYNDP scenario. As a result, we define 22 scenarios for 2030. We assume that these scenarios have equal probabilities of realisation. Figure \ref{figure_scenarios} (left) illustrates the construction of these 22 scenarios. We inter- and extrapolate each pair of scenarios by weighting them with factors of -10\%, 33\%, 50\%, 67\% and 110\%, thus obtaining five new scenarios. The interpolation between a TYNDP scenario and the EV is conducted using the mean of the two scenarios.

As mentioned before, our model horizon contains three representative years: 2020, 2025 and 2030. The year 2020 is assumed to be subject to no uncertainty. Hence, all scenario paths begin with the same assumptions. These are based on the `best estimate’ scenario, which also is provided by the TYNDP and represents an intermediate step to the scenarios introduced. The uncertainty regarding the four key parameters thus only occurs from 2025 onwards. The investment decisions are made in the first stage and must hold for all scenario realisations in the second stage. The scenario paths with the highest (endogenous) operational costs form the CVaR and drive the investments of risk-averse decision-makers.

\subsection{Data}
\label{subsec:data}

As mentioned above, scenario-specific assumptions regarding the uncertain input parameters of electricity demand, installed capacities of RES and the prices of fuel and CO\textsubscript{2} are defined by the TYNDP report from \textcite{tyndp2018}. A detailed overview of the scenario data is provided in Appendix A.

Fuel prices that are not scenario-specific (e.g. prices for nuclear power plants) are based on \textcite{Schröder:2013}. Node-specific thermal and hydro generation capacity, efficiency and decommission pathways are derived from \textcite{Schröder:2013}, \textcite{Gerbaulet2017}, \textcite{EU2016} and \textcite{OPSD}.

Investment costs for new electricity generation and storage capacities are derived from \textcite{Schröder:2013}, and investment costs for transfer capacities are based on \textcite{Gerbaulet2017}. Since we identify optimal investment decisions using three representative years, we compute annualised capital costs and allocate them to the respective year using an interest rate of 6\%. Lifetime assumptions regarding electricity generation and storage technologies are provided by \textcite{Schröder:2013} and can be found in the supplementary material (https://github.com/BTU-EnerEcon/RiskAv). For transfer capacities, we assume a lifetime of 50 years.

In reality, minimum capacity requirements ensure that new power plants are economically viable; our model, however, excludes these requirements. Instead, the model allows investment in marginal amounts of generation capacity, which are grouped into capacity clusters. However, we consider political and technical constraints on investments in generation capacities (e.g. the installation of nuclear, lignite or hard coal plants is only possible in countries without phase-out plans). Given the long planning horizon for most generation technologies, we assume that a capacity expansion in 2020 is feasible only for open cycle gas turbines. Run-of-river stations, hydroelectric reservoirs, biomass and waste-to-energy are not subject to an endogenous investment decision and follow a pre-defined expansion plan derived from \textcite{EU2016}. 

As storage technology, we consider only hydroelectric PSPs. These are modelled with their turbine and pumping capacity as well as their storage capacity. We connect these components assuming a capacity-power factor of nine, which is defined as the number of hours a fully charged storage system can produce at full load. Generation by hydroelectric reservoirs is constrained by the installed capacities and a monthly water budget. To determine the water budget, we use empirical electricity generation data; both empirical electricity generation data and installed capacities are taken from \textcite{EntsoeTP}.

Electricity generation from intermittent renewable capacities is not dispatchable and depends on meteorological conditions. We implement hourly feed-in profiles derived from \textcite{EntsoeTP} for onshore wind, offshore wind and photovoltaics. Despite the existence of different `wind-years', we assume that the hourly feed-in profiles do not vary within our model horizon. For the years considered (including their scenario variations), we adjust only the capacity levels. 

We allow for endogenous cross-border trade within the model's geographical scope. Electric power transmission between nodes is restricted by NTCs. We apply current NTC values for the year 2020 and implement the currently planned expansions until 2030. Hence, the endogenous investments in transfer capacities are always additional to the current expansion plan. Both current installations and currently planned expansions are taken from \textcite{tyndp2018}. We neglect intranational limitations on electricity flows. 

We model demand response with a location-specific supply function for demand reduction. Demand response activities, which are driven by a scarcity of power plant capacity, are assumed to be remunerated by the value of lack of adequacy (VoLA), which is individually determined for each industry sector and each European country by \textcite{CEPA:2018}. For each country, we identify the sector-specific share of electricity consumption \cite{EUROSTAT:2020} and link it to the hourly demand, constructing a stepwise supply function for load shedding activities. The resulting values are made available in the supplementary material (https://github.com/BTU-EnerEcon/RiskAv). When we apply these supply functions, we implicitly assume that each sector can perfectly signal its value of electricity to the market, such that if the costs of meeting (sectoral) demand exceed the (sectoral) valuation of electricity, the market outcomes include a reduction in demand. In what follows, we use the terms demand reduction' and `load shedding' as synonyms since all load shedding in our model is planned rather than a result of unexpected contingencies. Note that while domestic electricity consumption is included as one of the sectors, it is unlikely to be called upon since it has a much higher VoLA than most others. 

Since our optimisation model minimises the present value of the electricity system's total cost, all future costs are discounted. We apply a discount rate of 6\% per year.

\section{Results}
\label{sec:results}
This section is structured as follows. First, we illustrate the impact of risk aversion on investments in our base case, with all flexibility options (demand response, international trade and storage) at their current levels. Second, we vary the availability and cost of individual flexibility options to examine their impact on the risk mitigation measures taken by risk-averse planners. Finally, we analyse the effects of the interplay between flexibility elements.

\subsection{Base case}
\label{subsec:results_section_1}

\begin{result}
Risk-averse decision-makers consider the combined effect of all uncertainties. Since uncertainties are correlated and have different levels of impact, this can lead to counterintuitive effects. In our case, risk aversion has little impact on overall generation investment despite the presence of demand uncertainty. This effect occurs because uncertain CO\textsubscript{2} prices have a larger impact on investment decisions than other uncertainties, and CO\textsubscript{2} prices correlate negatively with demand levels in the framework of the TYNDP scenarios.
\end{result}

Risk-averse investment planning aims to avoid the very worst outcomes. Hence, investment decisions in a system with risk-averse agents differ from those made under risk-neutral decision-making. Assuming a hypothetical naïve approach in which a risk-averse planner addresses each uncertainty individually, one can derive the following principles for the four key uncertainties considered in our paper:
\begin{enumerate}
        \item Demand uncertainty would be addressed by installing more peak-load capacity, reducing costly load shedding in times of demand spikes.
        \item Like demand uncertainty, uncertainty about RES capacities would be addressed by installing more peak-load capacity, reducing the costly use of demand flexibility in times of low RES feed-in.
        \item CO\textsubscript{2} price uncertainty would be addressed by favouring investments in low-carbon technologies.
        \item Fuel price uncertainty would be addressed by favouring investments in technologies with lower price uncertainty.
\end{enumerate}

Assuming that the naïve approach is applied to mitigate the combined risk of the four uncertain parameters, a risk-averse planner could be expected to reduce risk by installing more generation capacity consisting of more low-carbon generators and generators with less variation in fuel costs. 

\begin{figure}[H]
\includegraphics[width=\textwidth]{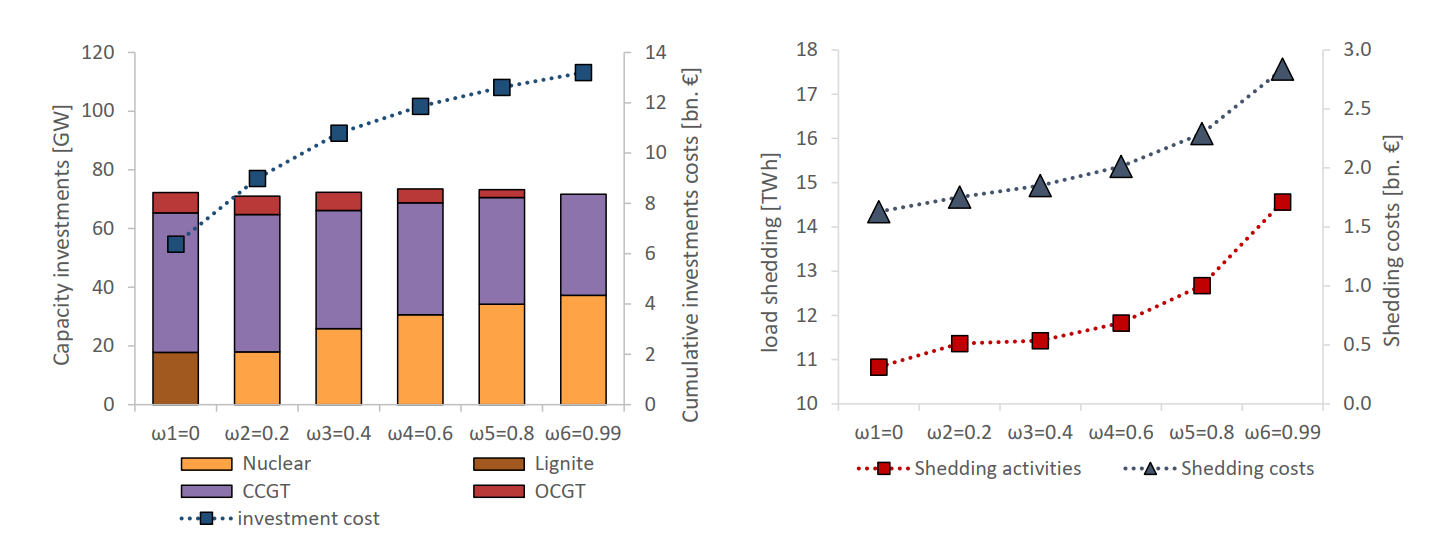}
	\caption{Aggregated European capacity investments until 2030 (left) and aggregated European load shedding activities and load shedding costs in 2030 (right) in risk-aversion scenario}
    	\label{figure_capacity_investments}
\end{figure}

Figure \ref{figure_capacity_investments} presents the aggregated European capacity investments through 2030 for different levels of risk aversion, from risk neutrality in $\omega \textsubscript{1}$ to very strong risk aversion in $\omega \textsubscript{6}$. Notably, risk-averse planning produces results within the framework of the TYNDP scenarios that deviate from the intuitions derived above. As risk aversion increases, we observe no increase in secured capacities, while aggregated European load shedding activities increase by roughly 30\% and the shedding costs increase by 70\%. This highlights that risk-averse system planners do not focus on installing more capacity. 

Furthermore, the investment mix has a clear trend in $\omega$. Figure \ref{figure_capacity_investments} shows that, in a risk-neutral setting, significant investments are made in lignite-fired generation capacity. With increasing risk aversion, lignite disappears from the energy mix, while nuclear technology enters it. This highlights the significant influence of CO\textsubscript{2} price uncertainty on decisions made by risk-averse planners. The risk of very high CO\textsubscript{2} prices pushes risk-averse planners to invest in more capital-intensive but emission-free nuclear power plants. The trend depicted in Figure \ref{figure_capacity_investments} shows that higher risk aversion corresponds to more emission-free nuclear capacity. In addition, we observe no open cycle gas turbine (OCGT) investment in $\omega \textsubscript{6}$. This can be explained by the scenarios with high CO\textsubscript{2} prices, which receive a very high weight if decision-makers are extremely risk averse, and the relatively low efficiency of the OCGT technology.

\begin{figure}[H]
\includegraphics[width=\textwidth]{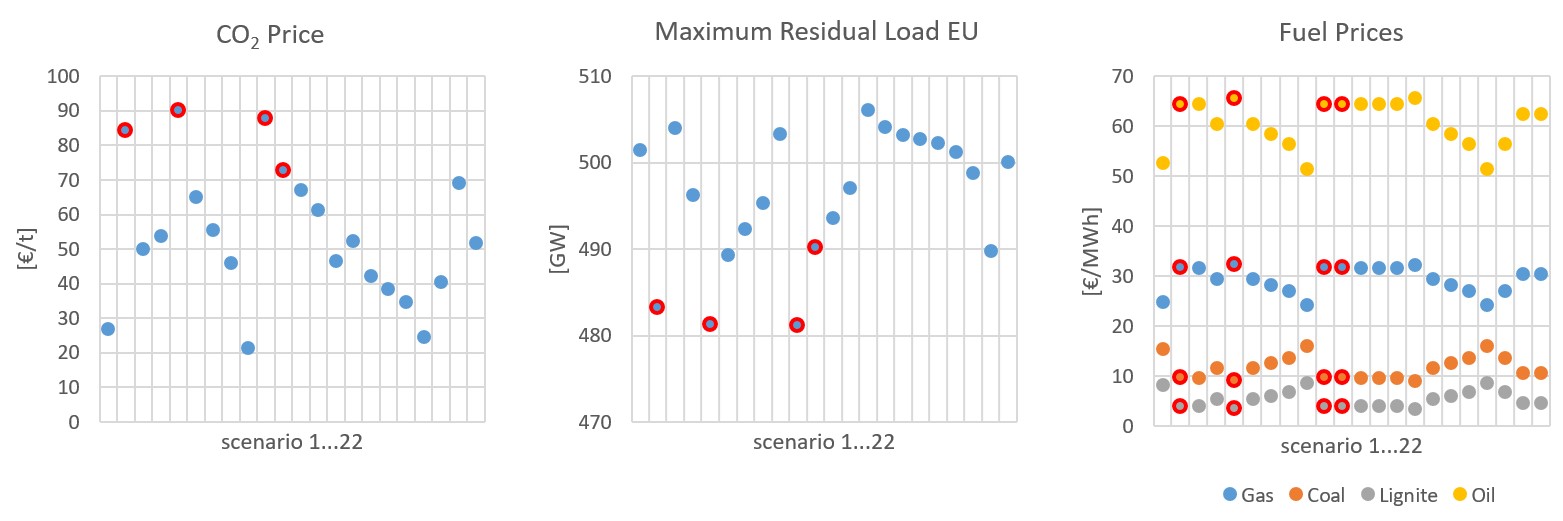}
	\caption{Scenario-specific CO\textsubscript{2} price (left), maximum hourly residual load aggregated for Europe (centre) and fuel prices (right) for the year 2030. The dots marked with red circles depict the scenarios that the risk-averse planner considers particularly risky.}
    	\label{figure_CO2_Dem_scenario}
\end{figure}

To further examine these results, we plot the parametric assumptions of each of our 22 scenarios in Figure \ref{figure_CO2_Dem_scenario}. Note that the demand and RES deployment scenarios can be illustrated together by plotting maximum hourly residual values. The scenarios that risk-averse planners consider as particularly risky and/or difficult to mitigate (i.e. those that are endogenously included in the CVaR tail and receive a higher weight) are highlighted.

As Figure \ref{figure_CO2_Dem_scenario} shows, the tail scenarios are exclusively the scenarios with the highest CO\textsubscript{2} prices. Because the TYNDP scenarios are internally consistent, the highlighted scenarios also have the lowest maximum residual loads, which means that they do not require extensive capacity investments to meet demand. Moreover, in the tail scenarios, lignite and hard coal prices tend to be low, while gas and oil prices tend to be high. Despite low lignite prices, risk-averse planners prefer to invest in nuclear capacity to hedge against the risk of high CO\textsubscript{2} prices. 

Thus, the interplay of parameters leads to (i) more investment in a less emission-intensive energy system if planners are risk averse (hedging against CO\textsubscript{2} price uncertainty) and (ii) constant total installed capacity, regardless of the level of risk aversion (planners do not hedge against demand and RES deployment uncertainties). 

\subsection{Risk aversion and flexibility elements}
\label{subsec:Impact of flexibility elements} 

Usually, more flexibility in electricity markets reduces investment risk, since flexibility options both reduce short-term price variability within scenarios and allow the electricity system to more efficiently adapt to long-term uncertainty between scenarios. Therefore, we would expect the availability of flexibility options to reduce the cost of risk mitigation in risk-averse system planning, thus reducing the effects of risk aversion. In this section, we investigate the impact of flexibility elements on a system in which decision-makers are endowed with different levels of risk aversion. We focus our discussion on three key flexibility elements of electricity markets: demand response, cross-border trade and energy storage. 

Each flexibility element has its own mechanism and dynamics. Demand response allows consumers to react to price spikes by reducing demand, receiving the sector-specific load shedding supply cost (VoLA\footnote{The value of lack of adequacy (VoLA) is described in section \ref{subsec:data}}) in return. Cross-border trade enables spatial transfers of electricity between regions with price differences and is limited by the availability of NTCs. Energy storage enables temporal transfers between periods with different prices. 

In sections \ref{subsubsec:demand response}--\ref{subsubsec:storage}, we examine the effects of the individual flexibility elements in isolation. Finally, in section \ref{subsubsec:interplay} we analyse the interplay of the three flexibility elements. 

In our discussion, we refer to the ex-post expectation of total system costs, which is the sum of all first-stage investment costs for generation, storage and cross-border transmission capacity expansion, plus a risk-neutral expectation of the second-stage operating costs. We use this metric instead of the ex-ante risk-averse expectation of total costs because the latter, by definition, increases with risk aversion since higher weights are placed on the costliest scenarios. Therefore, it would be difficult to disentangle the effects on total costs of a change in first-stage decisions from the effects of a simple change in scenario weights. Using a risk-neutral ex-post expectation isolates the effect on costs of a change in investment. Nevertheless, it is important to keep in mind that this metric is a mathematical construct and not the model objective. This means that, unlike the model objective, it can decrease even if we give the model more flexibility options to invest in.

\subsubsection{The isolated effect of demand response}
\label{subsubsec:demand response} 
\begin{result}
The value of demand flexibility generally increases with higher levels of risk aversion, implying that demand response has a risk-reduction value in addition to its expected-value efficiency benefit.
\end{result}
\begin{result}
Risk aversion has the largest impact on the overall system for low and intermediate levels of demand flexibility. If demand is highly flexible, the overall need for capacity -- and consequently the investment risk -- is reduced, so levels of risk are also low. If demand is inflexible, investment is driven purely by peak-load hours rather than by uncertain operating costs, so risk aversion also has a smaller effect on optimal decisions.
\end{result}

In this section, we investigate the impact of demand response on an electricity system with different levels of risk aversion. Aiming to isolate the effect of demand responsiveness, we create a European merit order of load shedding potential. As described above, we (i) average the sector-specific costs provided by \cite{CEPA:2018} and (ii) couple the resulting load shedding cost with sector-specific shares of electricity consumption for each country.

In this analysis, we model four settings:
    \begin{enumerate}
        \item \emph{No demand responsiveness}: demand is inelastic and load shedding is not possible.
        \item \emph{Low demand responsiveness}: load shedding is possible at high costs. The cost values of the merit order for load shedding are increased by a factor of five.
        \item \emph{Intermediate demand responsiveness}: load shedding is possible at estimated costs. The cost values of the merit order for load shedding are not scaled.
        \item \emph{High demand responsiveness}: load shedding is possible at low costs. The cost values of the merit order for load shedding are halved.
    \end{enumerate}

In Table \ref{table_TC_demandelasticity}, we list the defined model settings and show the resulting ex-post expected total system costs and European load shedding activities. Naturally, a system with no demand responsiveness has the highest ex-post expected total system cost. When demand responsiveness increases from low to high, we observe decreasing costs. The differences quantify the value of the additional demand responsiveness. For all four settings, ex-post expected total system costs increase with risk aversion. This increase is a result of the nonzero cost of risk mitigation. 

Table \ref{table_TC_demandelasticity} also lists the aggregated quantities of European load shedding activities. In the `no demand responsiveness' setting, shedding does not occur and the system invests in enough generation capacity to satisfy peak net demand in all scenarios. With increasing demand responsiveness, more shedding occurs. 
However, the increasing load shedding does not lead to certain sectors being permanently shut down. At its maximum, the proportion of hours in a year in which load shedding occurs ranges from 1.7\% in Sweden and Norway to 17.4\% in the UK.

Note also that shedding activities increase at higher risk-aversion levels -- that is, risk-averse decision-makers use the load shedding option to mitigate risk. However, the relative increase in load shedding activities from risk-neutral to risk-averse system planning differs across demand responsiveness settings. The increase is stronger for the low and intermediate levels (88\% and 84\%, respectively) compared to high demand responsiveness (30\%). This result illustrates that risk aversion has the largest impact on the energy system for intermediate levels of demand flexibility.

\begin{table}[H]
	\begin{center}
	\bigskip
	\caption{Ex-post expectation of total system costs at different levels of demand elasticity in bn. € and aggregated European load shedding activities in GWh}
    	\label{table_TC_demandelasticity}
    	\small
		\begin{tabular}{|m{10em}|r|r|r|r|r|r|}
			\hline
			\multicolumn{1}{|c|}{} &
			\multicolumn{1}{c|}{$\omega \textsubscript{1}$=0} &
			\multicolumn{1}{c|}{$\omega \textsubscript{2}$=0.2} &
			\multicolumn{1}{c|}{$\omega \textsubscript{3}$=0.4} &
			\multicolumn{1}{c|}{$\omega \textsubscript{4}$=0.6} &
			\multicolumn{1}{c|}{$\omega \textsubscript{5}$=0.8} &
			\multicolumn{1}{c|}{$\omega \textsubscript{6}$=0.99}\\
			\hline
			[1] no demand responsiveness & 186.9 & 187.2 & 187.3 & 187.3 & 187.5 & 189.4 \\
			\hline
			[2] low demand responsiveness & 186.4 & 186.6 & 186.7 & 186.8 & 186.9 & 188.6\\
			\hline
			[3] intermediate demand responsiveness & 184.3 & 184.5 & 184.6 & 184.8 & 186.8 & 187.0 \\
			\hline
			[4] high demand responsiveness & 183.2 & 183.4 & 183.5 & 183.6 & 183.8 & 185.6 \\
			\hline
			Load shedding at [1] & - & - & - & - & - & -  \\
			\hline
			Load shedding at [2] & 563 & 657 & 810 & 908 & 1,005 & 1,059 \\
			\hline
			Load shedding at [3]  & 6,031 & 6,703 & 7,191 & 8,144 & 8,851 & 11,084 \\
			\hline
			Load shedding at [4] & 17,200 & 17,954 & 18,399 & 18,949 & 19,903 & 22,277 \\
			\hline

		\end{tabular}
	\end{center}
\end{table}

\subsubsection{The isolated effect of trade}
\label{subsubsec:trade} 
\begin{result}
Cross-border transmission capacity expansion has a higher value for risk mitigation when decision-makers are risk averse. 
\end{result}

\begin{result}
Investment in cross-border capacity additional to the ENTSO-E TYNDP expansion plan has a negligible impact on the ex-post expected total system costs. This result, at least from the perspective of energy trading, indicates that the existing system, in combination with the currently planned expansions, can theoretically achieve most of the efficiency gains of cross-border trading if this trading is frictionless.
\end{result}

In this section, we investigate the impact of trade within an energy system with different levels of risk aversion. Hourly demand and renewable generation patterns vary between European countries. In times of low residual load (i.e. when supply is not scarce), free production capacities can be exported to neighbouring markets. In times of high residual load (i.e. when supply is scarce), electricity can be imported from neighbouring markets. Thus, trade ensures the efficient utilisation of generation capacities. However, both importing and exporting electricity require available transmission capacity between markets. Hence, increasing transmission capacity increases system flexibility. 

In this analysis, we model three settings:
    \begin{enumerate}
        \item \emph{No endogenous NTC expansion}: The TYNDP expansion plan is implemented in full, but further endogenous transmission expansion is not possible.
        \item \emph{Endogenous NTC expansion}: Further endogenous transmission expansion is possible at the current CAPEX estimate.
        \item \emph{Endogenous NTC expansion at reduced costs}: Further endogenous transmission expansion is possible at 50\% of the current CAPEX estimate. 
    \end{enumerate}

Note that the TYNDP expansion plan for 2030 is implemented exogenously in all three settings. Hence, the NTC investments derived from the model are to be understood as additional to the current expansion plan. For example, the 50\% CAPEX assumption in the third setting can be seen as the result of political efforts to intensively interconnect European energy markets.

Table \ref{table_TC_trade} depicts the ex-post expected total system costs and selected transmission capacities. It reveals several insights into the role of cross-border energy trade in risk-averse planning.

First, although the value of additional investments in cross-border capacity is non-negative in our model setup, we find that the actual changes in ex-post expected total system costs are small (columns in Table \ref{table_TC_trade}). This result indicates that the currently planned system can theoretically achieve most of the efficiency gains of cross-border trading, assuming that this trading is frictionless. Naturally, cross-border transmission has additional benefits beyond straightforward arbitrage, which we do not consider here.

Second, both the ex-post expected total system costs and the investment quantities increase with higher levels of risk aversion in all three settings (rows in Table \ref{table_TC_trade}). The positive correlation between invested capacity and risk-aversion levels indicates that flexibility from transmission expansion has a higher value for risk-averse system planning; its deployment for risk mitigation increases with risk aversion.

Overall, an interconnected energy system aims to exploit synergies between the residual load patterns in different nodes and to optimise the utilisation of generation capacities. Since risk-averse planning does not address uncertainty in peak demand by increasing installed generation capacities (see \ref{subsec:results_section_1}), efficient capacity utilisation is especially important. This drives the need for additional transmission capacity, which strengthens the interconnection across Europe. A detailed list of newly build interconnectors is presented in Appendix C. 

\begin{table}[H]
	\begin{center}
	\bigskip
	\caption{Ex-post expectation of total system costs with and without the possibility of trading electricity and building interconnectors in bn. € and endogenous investments in transfer capacities in MW}
    	\label{table_TC_trade}
    	\small
		\begin{tabular}{|m{12em}|r|r|r|r|r|r|}
			\hline
			\multicolumn{1}{|c|}{} &
			\multicolumn{1}{c|}{$\omega \textsubscript{1}$=0} &
			\multicolumn{1}{c|}{$\omega \textsubscript{2}$=0.2} &
			\multicolumn{1}{c|}{$\omega \textsubscript{3}$=0.4} &
			\multicolumn{1}{c|}{$\omega \textsubscript{4}$=0.6} &
			\multicolumn{1}{c|}{$\omega \textsubscript{5}$=0.8} &
			\multicolumn{1}{c|}{$\omega \textsubscript{6}$=0.99}\\
			\hline
			[1] no endogenous NTC expansion & 186.9 & 187.2 & 187.3 & 187.3 & 187.5 & 189.4 \\
			\hline
			[2] endogenous NTC expansion & 186.9 & 187.1 & 187.2 & 187.4 & 187.7 & 188.4 \\
			\hline
			[3] endogenous NTC expansion at reduced costs & 186.7 & 187.0 & 187.1 & 187.3 & 187.3 & 188.0  \\
			\hline
			Endogenous NTC investments at [2] & 6,459 & 7,447 & 8,763 & 10,214 & 14,586 & 15,601 \\
			\hline
			Endogenous NTC investments at [3] & 15,479 & 15,107 & 17,021 & 24,402 & 26,086 & 27,789 \\
			\hline
		\end{tabular}
	\end{center}
\end{table}

\subsubsection{The isolated effect of energy storage}
\label{subsubsec:storage} 
\begin{result}
Like demand response and cross-border trading, the value of storage increases with the level of risk aversion in the market.
\end{result}

In this section, we investigate the impact of energy storage on systems with different risk-aversion levels. 

Storage units withdrawal electricity from the system at a certain point in time and re-inject it at any later point, reduced by efficiency losses. Thus, storage provides the flexibility needed to reduce imbalances between supply and demand. While energy storage technology has various forms, we focus exclusively on pumped-storage plants (PSPs) to avoid unnecessary complexity.

In this analysis, we model three settings:
    \begin{enumerate}
        \item \emph{No endogenous PSP expansion}: Endogenous storage expansion is not possible.
        \item \emph{Endogenous PSP expansion}: Endogenous storage expansion is possible at the current CAPEX estimate.
        \item \emph{Endogenous PSP expansion at reduced costs}: Endogenous storage expansion is possible at 50\% of the current CAPEX estimate. 
    \end{enumerate}

Note that the existing storage capacities are present in all three settings. Hence, PSP investments derived from the model are to be understood as an extension of currently installed capacities. Similar to the assumption made in section \ref{subsubsec:trade}, a 50\% CAPEX setting could be interpreted as a future in which an increased political effort is made to promote storage investment in Europe; it could also simply be a result of unexpected technological progress.

Table \ref{table_storage} reports the ex-post expected total system costs and realised storage investments. First, we find a minor decrease in ex-post expected total system costs as the availability of storage investments increases (columns in Table \ref{table_storage}). 

Second, we observe no storage investments at current cost levels (row $[2]$), which means that the price spread between base and peak times does not compensate for additional storage investments. Note, however, that we do not consider any revenue streams for storage except for the energy-only market. In the reduced CAPEX scenario (row $[3]$), we observe significant storage investments.

Third, the quantity of storage investments follows an interesting pattern under different levels of risk aversion. Storage investments remain constant for the first three steps ($\omega\textsubscript{1}--\omega\textsubscript{3}$); the investments increase at higher values of risk aversion. At lower levels of risk aversion (and indeed, in the risk-neutral setting), the entire potential storage capacity in France and the Iberian Peninsula is exploited; these regions have high price spreads because of the composition of the local generation mix. Only if risk aversion is high enough to overcome the significant costs of PSPs further investments are made in Belgium.

\begin{table}[H]
	\begin{center}
	\bigskip
    \caption{Ex-post total system costs with and without the option to invest in storage in bn. € and PSP investments in MW}
    	\label{table_storage}
    	\small
		\begin{tabular}{|m{11em}|r|r|r|r|r|r|}
			\hline
			\multicolumn{1}{|c|}{} &
			\multicolumn{1}{c|}{$\omega \textsubscript{1}$=0} &
			\multicolumn{1}{c|}{$\omega \textsubscript{2}$=0.2} &
			\multicolumn{1}{c|}{$\omega \textsubscript{3}$=0.4} &
			\multicolumn{1}{c|}{$\omega \textsubscript{4}$=0.6} &
			\multicolumn{1}{c|}{$\omega \textsubscript{5}$=0.8} &
			\multicolumn{1}{c|}{$\omega \textsubscript{6}$=0.99}\\
			\hline
			[1] no endogenous PSP expansion & 186.9 & 187.2 & 187.3 & 187.3 & 187.5 & 189.4 \\
			\hline
			[2] endogenous PSP expansion & 186.9 & 187.2 & 187.3 & 187.3 & 187.4 & 189.1 \\
			\hline
			[3] endogenous PSP expansion at reduced costs & 186.8 & 187.0 & 187.1 & 187.2 & 187.4 & 188.1 \\
			\hline
            PSP investments at [2] & - & - & - & - & - & - \\
			\hline
			PSP investments at [3] & 15,964 & 15,964 & 15,964 & 15,964 & 16,358 & 17,256 \\
			\hline
		\end{tabular}
	\end{center}
\end{table}

\subsubsection{Interplay of flexibility elements}  
\label{subsubsec:interplay} 

\begin{result}
Different sources of flexibility can be complements or substitutes depending on the level of risk aversion. 
\end{result}

In the following section, we analyse the interplay of flexibility elements under different risk-aversion levels. This is done by including all three flexibility elements in the model's formulation. In this analysis, we apply two settings that differ with respect to system flexibility:
    \begin{enumerate}
        \item \emph{Moderately flexible system}: includes the European merit order for load shedding, as well as the options to invest in NTCs and storage at current cost levels.
        \item \emph{Highly flexible system}: replicates the setting above, but cost levels are reduced by 50\%. Thus, we combine `flexible' settings from sections \ref{subsubsec:demand response}--\ref{subsubsec:storage}.
    \end{enumerate}

Table \ref{table_interplay_utilization_intermediate_flexible} shows the results for the \emph{moderately flexible system} setting. This table depicts the utilisation of the three flexibility elements under different risk-aversion levels (rows 1--3) and the differences between these values and the values observed in sections that consider respective flexibility elements in isolation (\ref{subsubsec:demand response}--\ref{subsubsec:storage}, rows 4--6). Hence, a negative value in the `Delta load shedding [GWh]' row indicates a decrease in load shedding in a setting if all three elements are available. Conversely, a positive value indicates an increase. Similarly, Table \ref{table_interplay_utilization_very_flexible} presents the results for the \emph{highly flexible system} setting.

In the \emph{moderately flexible system}, load shedding activities and NTC investments increase with higher risk aversion, but storage facilities are not built. However, the delta rows show small differences in utilisation for the flexibility elements (i.e. load shedding volumes and investments in NTCs and storage) when compared to settings in which each element is implemented in isolation. Hence, although the different flexibility elements are substitutes or complements at different levels of risk aversion, these effects are very small and exhibit no clear patterns

In the \emph{highly flexible system}, the picture is different. Load shedding activities increase moderately at higher values of risk aversion. The investments in NTCs increase remarkably from 19.3 to 53.7 GW. In contrast, the investments in PSPs decline from 1.5 to 0 GW with higher risk aversion. When compared to model settings with isolated flexibility elements, this analysis shows significant differences. In particular, (i) PSP investments are significantly lower, (ii) load shedding activities increase moderately at lower levels of risk aversion but considerably decline when risk aversion is high and (iii) investments in NTCs are always higher than in model settings with isolated flexibility elements and increase progressively with higher levels of risk aversion.

This result implies that transmission expansion becomes the dominant flexibility element when the three elements are modelled together, in particular when system planners are more risk averse. In such a setting, storage units are no longer an efficient option to mitigate risk, independently of the risk preference. In general, flexibility options can be complements at low levels of risk aversion and substitutes at high levels, as is the case with energy storage here.

\begin{table}[H]
	\begin{center}
%	\bigskip
	\caption{Utilisation of flexibility elements as the difference between a moderately flexible system and the system with the respective isolated element in \ref{subsubsec:demand response} -- \ref{subsubsec:storage}}
    	\label{table_interplay_utilization_intermediate_flexible}
    	\small
		\begin{tabular}{|m{13em}|r|r|r|r|r|r|}
			\hline
			\multicolumn{1}{|c|}{} &
			\multicolumn{1}{c|}{$\omega \textsubscript{1}$=0} &
			\multicolumn{1}{c|}{$\omega \textsubscript{2}$=0.2} &
			\multicolumn{1}{c|}{$\omega \textsubscript{3}$=0.4} &
			\multicolumn{1}{c|}{$\omega \textsubscript{4}$=0.6} &
			\multicolumn{1}{c|}{$\omega \textsubscript{5}$=0.8} &
			\multicolumn{1}{c|}{$\omega \textsubscript{6}$=0.99}\\
    		\hline
    		Aggregated load shedding [GWh] & 5,946  & 6,701 & 7,165 & 8,093 & 8,884 & 11,299    \\
    		\hline
            Endogenous NTC investments [MW] & 6,722 & 7,521 & 8,925 & 9,813 & 11,866 & 15,639    \\
    		\hline
    		Endogenous PSP investments [MW] & - & - & - & - & - & -  \\
    		\hline
    		Delta load shedding [GWh] & -85 & -1 & -26 & -51 & 32 & 215 \\
    		\hline
    		Delta NTC investments [MW]  & 263 & 74 & 162 & -401 & -2,720 & 38 \\
    		\hline
    		Delta PSP investments [MW] & - & - & - & - & - & -  \\
    		\hline    		
    		
		\end{tabular}
	\end{center}
\end{table}

\begin{table}[H]
	\begin{center}
%	\bigskip
	\caption{Utilisation of flexibility elements as the difference between a highly flexible system and the system with the respective isolated element \ref{subsubsec:demand response} -- \ref{subsubsec:storage}}
    	\label{table_interplay_utilization_very_flexible}
    	\small
		\begin{tabular}{|m{13em}|r|r|r|r|r|r|}
			\hline
			\multicolumn{1}{|c|}{} &
			\multicolumn{1}{c|}{$\omega \textsubscript{1}$=0} &
			\multicolumn{1}{c|}{$\omega \textsubscript{2}$=0.2} &
			\multicolumn{1}{c|}{$\omega \textsubscript{3}$=0.4} &
			\multicolumn{1}{c|}{$\omega \textsubscript{4}$=0.6} &
			\multicolumn{1}{c|}{$\omega \textsubscript{5}$=0.8} &
			\multicolumn{1}{c|}{$\omega \textsubscript{6}$=0.99}\\
    		\hline
    		Aggregated load shedding [GWh] & 17,851 & 18,375 & 18,748 & 19,027 & 19,001 & 19,451 \\
    		\hline
            Endogenous NTC investments [MW] & 19,257 & 18,658 & 21,346 & 30.040 & 39,901 & 53,718 \\
    		\hline       		
    		Endogenous PSP investments [MW] &  1,476 & 2,026 & 1,571 & 594 & 98 & -  \\
    		\hline		
    		Delta load shedding [GWh] & 651 & 421 & 349 & 78 & -902 & -2,826 \\
    		\hline
    		Delta NTC investments [MW] & 3,778 & 3,551 & 4,325 & 5,638 & 13,815 & 25,929 \\
    		\hline
    	    Delta	PSP investments [MW] & -14,488 & -13,938 & -14,393 & -15,370 & -16,260 & -17,256 \\
    		\hline
		\end{tabular}
	\end{center}
\end{table}

%     \begin{itemize}
%        \item Assuming an equal VoLA for each node results in 'evenly' %distributed shedding. However, considering constant VoLA combined with %endogenous NTC investments, load shedding moves to north western Europe, to %countries like the UK, Netherlands and France. (Figure %\ref{figure_shed_node_eqVoLA})
%        \item Country-specific VoLA levels facilitate a concentration of %load shedding in countries with low shedding costs. For each country, Figure %7 depicts the cumulated load shedding activities in the year 2030 as a share %of the annual electricity demand. We observe bulks of load shedding %activities particularly in Slovakia, Romania, Sweden, Slovenia and Finland.
%    \end{itemize}    
    
%\begin{figure}
%\includegraphics[width=\textwidth]{images/shedding_VoLAequal_country-specific.JPG}
%	\caption{Shedding activities in 2030 with an average European VoLA at each node }
%    	\label{figure_shed_node_eqVoLA}
%\end{figure}

%\begin{figure}
%\includegraphics[width=\textwidth]{images/shedding_VoLA merit-order_country-specific.jpg}
%	\caption{Shedding activities in 2030 with country-specific VoLA at each node }
%   	\label{figure_shed_node_varVoLA}
%\end{figure}

\section{Discussion}
\label{sec:discussion}

The above results show several interesting effects. First of all, we find that risk aversion has a minimal effect on overall investment in electricity generation capacity but increases investments in low-carbon capacity. This is in line with the previous literature (e.g. \cite{Munoz2017}), but since we have taken renewable investment to be exogenous, the increase applies mostly to nuclear capacity. In particular, investing in nuclear generation capacity is not optimal under a strictly risk-neutral objective but is only selected if market participants are assumed to be risk averse. This finding shows that nuclear generation capacity has additional value in mitigating risk despite its high costs. It may also help to explain why countries such as the UK are actively pursuing new nuclear investments even though they appear to be a sub-optimal solution from a cost-based perspective. Finally, these results suggest that estimates of endogenous renewable investments in the previous literature have not unduly under- or overstated the effects of risk aversion.

The reason that low-carbon capacity is especially attractive to risk-averse planners is, of course, the fact that CO$_2$ price uncertainty is both the single largest uncertainty in our model and the costliest to hedge against. Again, this is consistent with previous findings. Our results illustrate that policymakers must consider risk aversion in order to accurately predict the impact of policy measures such as carbon markets. In addition, when sets of scenarios are being generated, considerable attention is normally paid to their internal consistency based on estimated or perceived correlations between uncertain parameters. This can cause unexpected results when these scenarios are used for, for example, long-term system planning. In particular, if high-impact uncertainties correlate negatively with low-impact uncertainties, low-impact uncertainties can appear to have the opposite impact on model outcomes than what their isolated impact would be. Any type of scenario-based work should consider this effect. In general, it is well-known that the outcomes of stochastic planning models depend to a great extent on the most extreme scenarios; in our case, the outcomes of risk-averse planning models depend to a great extent on the scenarios that are costliest to mitigate.

As expected, flexibility elements have more value in a risk-averse world; in other words, risk aversion has less impact in a flexible electricity system. This is good news for developers of flexible technologies because it implies that the risk-neutral models that have often been used to evaluate new investments have understated their benefits. Nonetheless, the results should also encourage more analysis of risk. 

However, flexibility options are not equal. Demand response, which can reduce demand, more effectively reduces risk than technologies that only shift demand in space or time, such as transmission or storage. Therefore, demand response has long-term risk-reduction value in addition to short-term flexibility and operational efficiency. In most current electricity markets, consumers have a very limited ability to signal their willingness and adjust their demand; our results provide all the more reason to change this.

This is particularly important in the medium term since there are (perhaps counterintuitively) nonlinearities in the relationship between flexibility and risk. In particular, the effects of risk are most pronounced at medium levels of demand flexibility. This occurs because, in a very inflexible system, generation investment is driven by the highest level of demand. As a result, the utilisation factors of peak-load capacity are low, and uncertainty about their marginal costs is less important. In a highly flexible system, less overall capacity is necessary, which also reduces risk. This implies that, in the energy transition, the coming decade will perhaps see the highest level of risk and the largest impact of risk aversion. This means that financial markets and other methods to mitigate or hedge risk are more important now than ever and that further investments in flexibility elements will have significant benefits.

Our results also show interesting interactions between flexibility elements. At the current cost levels for flexibility and for all risk-aversion levels, our results do not show synergies or substitution effects that significantly affect shedding activities or decisions about investment in transmission or storage capacity. However, this picture changes significantly when flexibility options become cheaper. Transmission expansion becomes the dominant flexibility element, replacing storage investments in particular, and this effect increases with higher risk aversion. This means that in future electricity markets, flexibility provided by efficient electricity trading will create more system value than ever. This value is realised through arbitrage opportunities across time and space to the extent permitted by the transmission infrastructure. This effect creates a system benefit of further integration of the current European energy system and supports additional research on improving transmission infrastructure. On the contrary, risk-averse system planners will most likely not make additional storage capacity investments if they are optimising arbitrage revenues alone. Despite political efforts, this could inhibit both the expansion of storage technologies and the development of storable energy carriers, such as hydrogen.

%Our results also show spatial effects across Europe. Markets with industry structures that imply high levels of demand flexibility can import capacity inadequacy from neighbouring markets, capitalising on their flexibility. As our results show, there is value in this, relative to a one-size-fits-all approach where demand flexibility is costed equally across Europe. However, our results do illustrate the importance of European collaboration on flexibility, including discussions on how the benefits of flexibility, as well as their costs, can be shared; how capacity adequacy should be defined in an interconnected European context, etc. 

Naturally, our approach has limitations. First, we use a linear model that assumes completeness and perfect competition in both electricity and financial markets and does not include some of the nonlinearities that are present in real-world power system operations. In reality, incomplete financial markets will lead to an inefficient distribution of risk among market participants, increasing the cost of uncertainty and the effects of risk aversion. Moreover, nonlinearities in power plant operations can increase the need for flexible generation capacity, possibly reducing the amount of nuclear energy in an optimal generation mix. 

Second, our operationalisation of demand response, which is based on published values of lost adequacy, is a very rough approximation of real demand response potential and cost. Likewise, our NTC-based approach to transmission and transmission investment only approximates real-world network constraints and investment opportunities. 

Third, we only consider the value of generation capacity and flexibility elements in the electricity market. Naturally, most of these technologies and options have value in other applications, including ancillary services markets and secure power systems operation. Indeed, the current storage capacity in Europe derives a large part of its revenue from non-electricity services. 

Finally, as in all stochastic risk-averse models, our results depend to a large extent on the scenarios we have used. Scenarios are subjective by definition  and any set of scenarios describes only a limited range of possible futures. Our study is based on the TYNDP scenarios, which are a widely consulted and used set of scenarios that we consider representative of the scenarios that actual decision-makers use; in a Bayesian sense, they are therefore one of the best subjective scenario sets. However, individual decision-makers will use different scenarios with different probabilities, which we cannot capture. 

All of this means that our quantitative results may vary with model assumptions and parameters. Nevertheless, we expect our qualitative results to carry over to a wide range of model formulations and, indeed, to the real world. In particular, our qualitative results demonstrate the interplay between investment risk and flexibility. This is clearly an important dimension of energy systems planning and modelling that should receive increased attention in future research.
\section{Conclusion}
\label{sec:conclusion}

In this paper, we investigate risk-averse decision-making in the context of long-term system planning. We develop a pan-European stochastic electricity market model that accounts for the risk-averse decision-making behaviour of its agents.  %The uncertainty in our model is derived from the widely used TYNDP scenarios, which consider four key parameters on the electricity system that are subject to high uncertainty: electricity demand, RES capacities, CO$_2$ prices and fuel prices. 
We apply a set of risk-aversion levels and analyse 1) the impact of risk aversion on investments and 2) the use of flexibility options including demand response, storage and interconnection in such a system. In doing so, we generalise the literature on risk-averse planning to a setting that much more closely resembles future, flexible electricity markets. We also generalise the literature on the value of flexibility to settings where planners are risk-averse.

We find that risk aversion decreases investment incentives, especially for high-emission technologies. However, since risk-averse decision-makers consider all uncertainties in combination such that these uncertainties correlate with each other and have a different level of impact, we observe counterintuitive effects. In particular, we show that risk aversion does not lead to an overall increase in capacity investment despite the presence of demand uncertainty. 

Although we implement a variety of parameters that are subject to uncertainty, we find that in the European context, carbon price uncertainty generates the highest investment risk. Hence, as risk aversion increases, we see a shift in the investment mix from emission-intensive to emission-free technologies. This may explain why some countries are actively pursuing new nuclear investments even though this appears to be sub-optimal from a cost-based perspective.

Furthermore, we find that flexible technologies have a higher value for risk-averse decision-makers -- that is, mitigating risk is easier in a flexible electricity system. This implies that models that use risk neutrality as a central assumption, which have often been used to evaluate investment decisions, have underestimated the benefits of system flexibility. However, the impacts of risk are nonlinear; we observe an especially large impact of risk aversion on model outcomes for intermediate levels of flexibility.

Finally, we investigate the interplay between the flexibility elements. At the current cost levels for flexibility and for all risk-aversion levels, our results do not reveal significant synergies or substitution effects among flexibility elements. However, once flexibility is available at lower cost, transmission expansion becomes the dominant flexibility element, replacing storage investments; this effect increases at higher levels of risk aversion.

Further research on this topic is necessary and could address the limitations discussed above. Our model is formulated as a linear cost minimisation problem and considers demand flexibility only with the value of lack of adequacy. Future studies could form a nonlinear welfare maximisation problem that allows a more detailed representation of demand elasticity. Research on the extent to which nonlinearities in power plant operation increase the need for flexibility could also provide new and interesting insights. Finally, further analysis could extend the time period considered in this study to map the transformation of the European energy system to 2050 and beyond.

%In our study, we apply the well recognized and widely used TYNDP scenarios. However, the consideration of different scenario setting as well as a different implementation of probabilities can be beneficial and could be investigated.

\section*{Acknowledgements}
A.H. van der Weijde received support from the Alexander von Humboldt Foundation through a Humboldt Fellowship. 
Thomas Möbius received support through the research project ProKoMo, which is funded by the German Federal Ministry of Economic Affairs.
Iegor Riepin received support through the research project Ariadne (Kopernikus project), which is funded by the German Federal Ministry of Education and Research.

{\renewcommand{\markboth}[2]{}}
\printbibliography
\newpage
\pagestyle{empty}

\section*{Appendix}
\label{sec:appendix}

\textbf{Appendix A: Scenario Data}

\begin{table}[H]
	\begin{center}
	\bigskip
	\caption{Projections for installed RES capacities in GW}
    	\label{Appendix A_1}
    	\small
\begin{figure}[H]
\includegraphics[width=\textwidth]{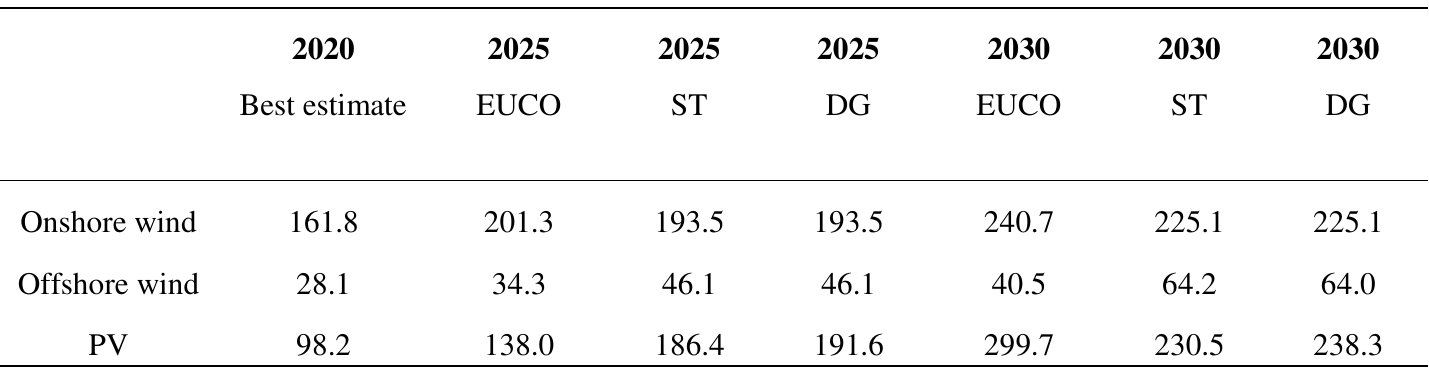}
\end{figure}		
	\end{center}
\end{table}

\begin{table}[H]
	\begin{center}
	\bigskip
	\caption{Fuel and CO$_2$  prices. Monetary values are in €${_2015}$}
    	\label{Appendix A_2}
    	\small
\begin{figure}[H]
\includegraphics[width=\textwidth]{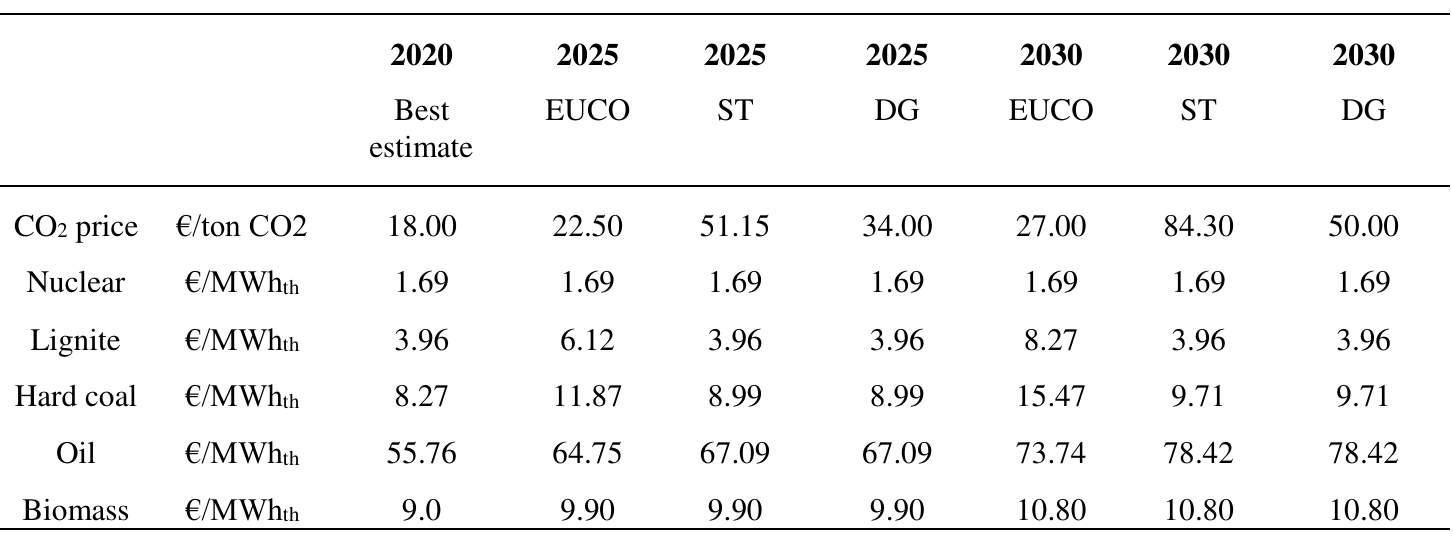}
\end{figure}		
	\end{center}
\end{table}

\begin{table}[H]
	\begin{center}
	\bigskip
	\caption{Characteristic demand levels in 2030 cumulative for all model regions in GWh}
    	\label{Appendix A_3}
    	\small
\begin{figure}[H]
\includegraphics[width=0.5\textwidth]{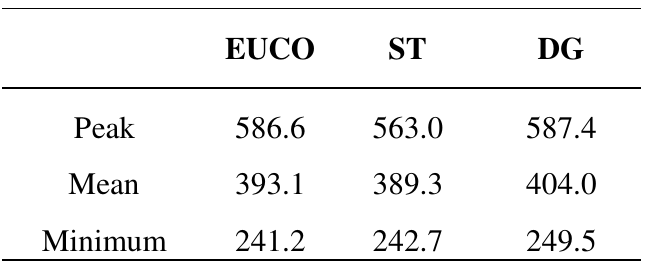}
\end{figure}		
	\end{center}
\end{table}

\newpage

\textbf{Appendix B: Geographical Scope}

Below, we list the countries and regions included in our model’s geographical scope.  
Note that the model structure consists of a network of nodes. A node represents one country or a group of several countries from one region. The geographical scope covers EU member states, European countries that are not part of the EU. We elicit node names as defined by the two-letter ISO 3166 international standard. In cases where countries are represented by a compound region with aggregated data, we set the node name in line with the name of the region.

\begin{table}[H]
	\begin{center}
	\bigskip
	\caption{Model regions}
    	\label{Appendix B_1}
    	\small
\begin{figure}[H]
\includegraphics[width=0.85\textwidth]{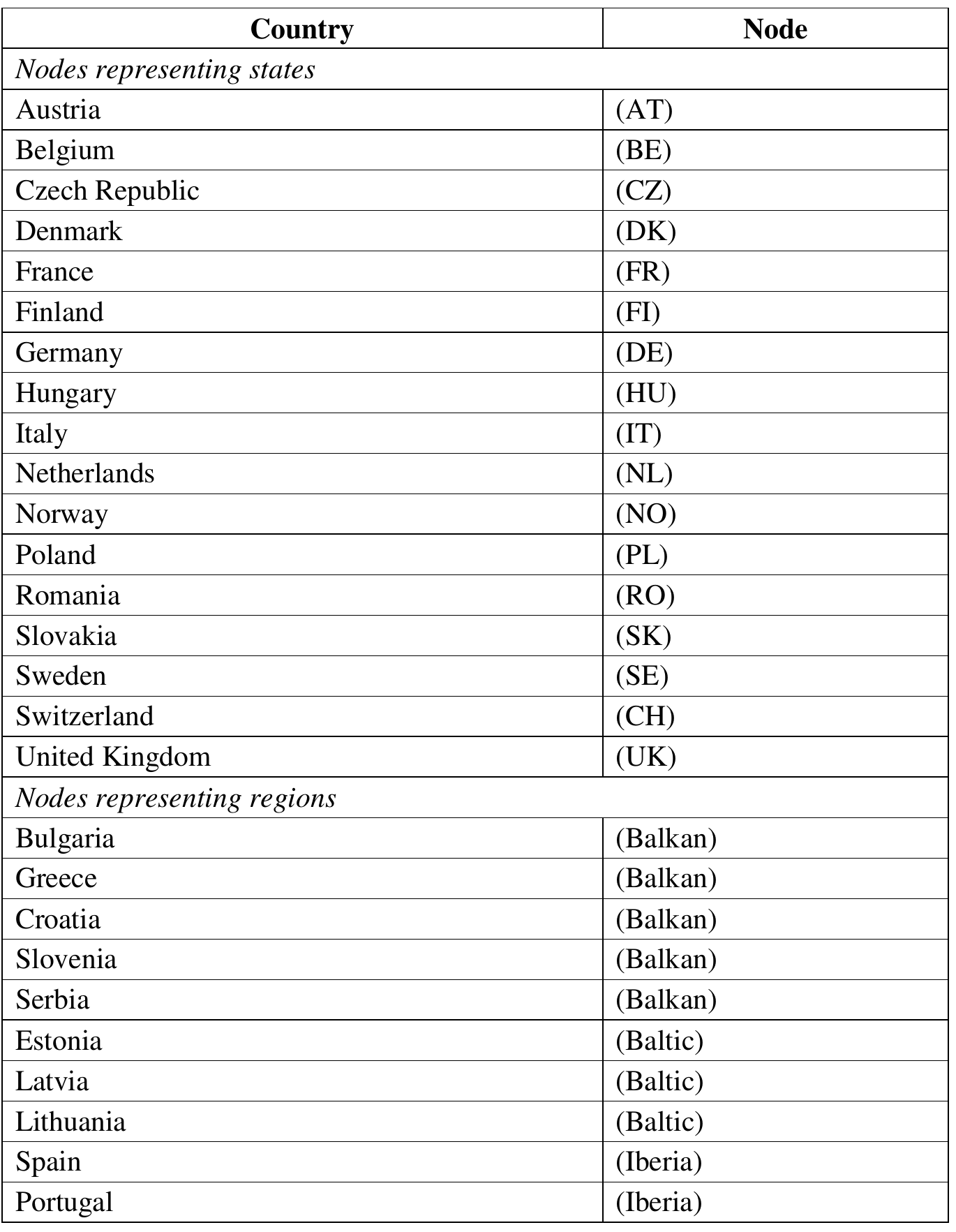}

\end{figure}		
	\end{center}
\end{table}

\newpage
\textbf{Appendix C: Supplementary Material}

\begin{table}[H]
	\begin{center}
	\bigskip
	\caption{Newly built NTCs in MW at currently estimated CAPEX}
    	\label{Appendix C_1}
    	\small
\begin{figure}[H]
\includegraphics[width=\textwidth]{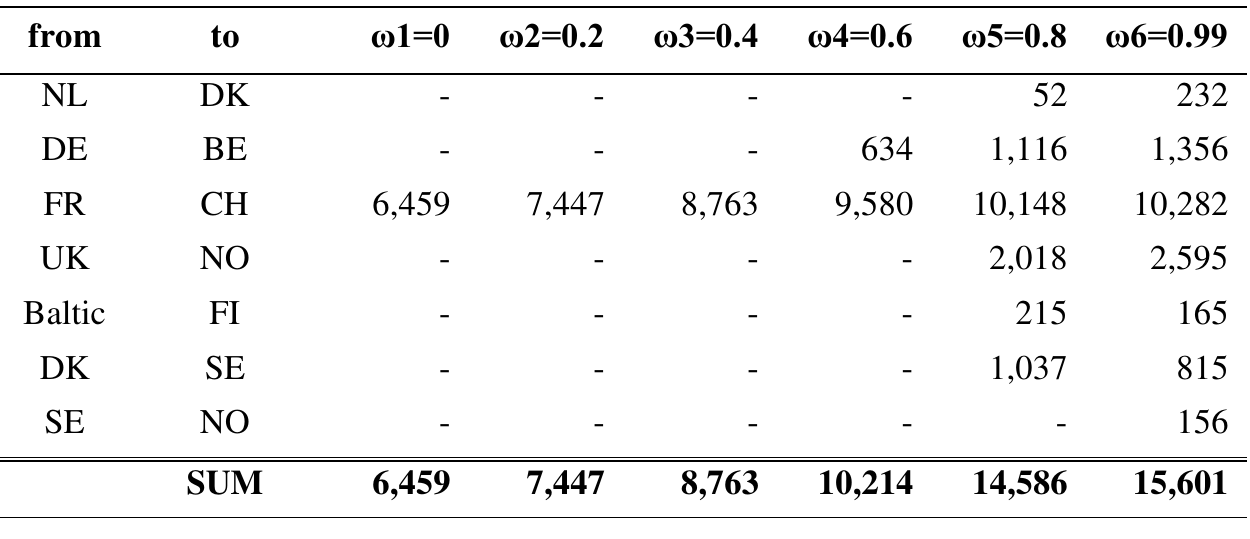}
\end{figure}		
	\end{center}
\end{table}

\begin{table}[H]
	\begin{center}
	\bigskip
	\caption{Newly built NTCs in MW at 50\% of current CAPEX}
    	\label{Appendix C_2}
    	\small
\begin{figure}[H]
\includegraphics[width=\textwidth]{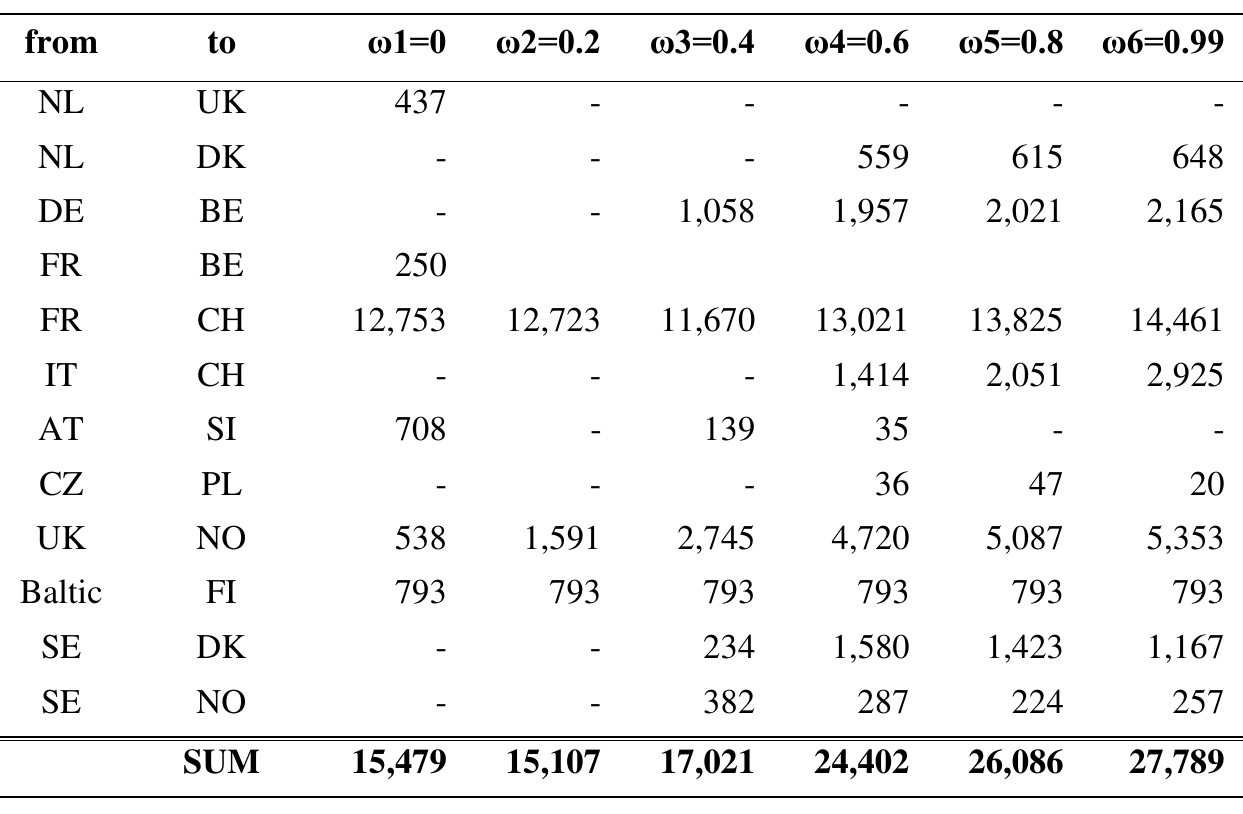}
\end{figure}		
	\end{center}
\end{table}
\end{document}